# Time-dependent level crossing models solvable in terms of the confluent Heun functions

**A.M. Ishkhanyan and A.E. Grigoryan**

Institute for Physical Research, NAS of Armenia, 0203 Ashtarak, Armenia

We discuss the level-crossing field configurations for which the quantum time-dependent two-state problem is solvable in terms of the confluent Heun functions. We show that these configurations belong to fifteen four-parametric families of models that generalize all the known 3- and 2-parametric families for which the problem is solvable in terms of the Gauss hypergeometric and the Kummer confluent hypergeometric functions. Analyzing the general case of variable Rabi frequency and frequency detuning we mention that the most notable features of the models provided by the derived classes are due to the extra constant term in the detuning modulation function. Due to this term the classes suggest numerous symmetric or asymmetric chirped pulses and a variety of models with two crossings of the frequency resonance. The latter models are generated by both real and complex transformations of the independent variable. In general, the resulting detuning functions are asymmetric, the asymmetry being controlled by the parameters of the detuning modulation function. In some cases, however, the asymmetry may be additionally caused by the amplitude modulation function. We present an example of the latter possibility and additionally mention a constant amplitude model with periodically repeated resonance-crossings. Finally, we discuss the excitation of a two-level atom by a pulse of Lorentzian shape with a detuning providing one or two crossings of the resonance. Using a series expansion of the solution of the confluent Heun equation in terms of the Kummer hypergeometric functions we derive particular closed form solutions of the two-state problem for this field configuration. The particular sets of the involved parameters for which these solutions are obtained define curves in the 3D space of the involved parameters belonging to the complete return spectrum of the considered two-state quantum system.

## 1. Introduction

The level-crossing is a key concept of the theory of non-adiabatic transitions [1] well appreciated for a long time starting from the pioneering works by Landau [2], Zener [3], Majorana [4], and Stückelberg [5]. Such models, both time-dependent and time-independent, have been widely studied in the context of many physical and chemical phenomena including, for example, magnetic resonance [2-5], electronic transitions in atomic and molecular collisions [6-9], laser-induced atomic dynamics [10-21], atom optics [22], atom lasers [23], laser cooling [24], dynamics of Bose-Einstein condensates [25], quantum dot molecule [26], molecular nanomagnets [27], quantum control and quantum Hall effect [28], nanophysics, quantum information processing with superconducting qubits [29], dynamics of quantum phase transitions [30], cold atom photo- and magneto- association into molecules [31], neutrino oscillations [32], chemical reactions [33,34], etc. Models dealing with level-crossings are also applied to study complex systems in biology [35,36].

In the present paper we discuss the level-crossing models for which the solution of the time-dependent two-level problem is written in terms of the confluent Heun function [37,38]. This function, the solution of the confluent Heun equation, is a member of the Heun class of mathematical functions that are believed to compose the next generation of special functions [37-40]. Among the five Heun equations generating these functions, the confluent Heun equation is of particular interest because it directly incorporates the hypergeometric and confluent hypergeometric equations, as well as the Mathieu equation [37-39]. Other known equations can be viewed as particular, transformed or limiting cases of this equation, e.g., the spheroidal, Coulomb spheroidal, generalized spheroidal wave equations, and the Whittaker-Hill equation [39]. For this reason, the analytic models solvable in terms of the confluent Heun function directly generalize many of the known solvable cases, particularly the ones solvable in terms of hypergeometric and confluent hypergeometric functions.

In total, there exist fifteen classes of models solvable in terms of the confluent Heun function. The classes are identified by a pair of functions referred to as basic integrable model. The first function referred to as the amplitude modulation function stands for the Rabi frequency of the field. The second function referred to as the detuning modulation function defines the time variation of the detuning. According to the general class property of the solvable two-state models [19-21], the actual field configurations are then generated, by applying different (in general, complex-valued) transformations of the dependent variable. Importantly, the confluent Heun functions provide extension of the previously known one- and two-parametric detuning modulation functions to the three-parametric case. This extension leads to models with non-linear sweeping through the resonance [13], chirped pulses with two time scales [14], level-glancing configurations [15], a variety of symmetric and asymmetric models of double passage through the resonance [16,17] and, in specific cases, periodically repeated crossing models [18].

A useful generic feature of the level-crossing models described by the confluent Heun equation is that they in general suggest, due to additional parameters involved in the Heun equation, processes with more time scales compared with the models described by the hypergeometric equations. For instance, in the case of chirped pulses we have a coupling that acts over a time interval which is not connected to the effective time of the resonance crossing. Another representative example is the case of double crossing models where the time-separation between two crossings and the speed the system crosses a particular resonance point are controlled almost independently, by separate parameters. The same advantageous feature is observed when periodically repeated crossings are discussed: the coupling strength, the period of crossings and the detuning modulation amplitude are described by well identified separate parameters. For this reason, below we discuss the possible crossing models and explore the basic properties of the corresponding field configurations in terms of the parameters of the confluent Heun equation.

**2. Fifteen classes of models solvable in terms of the confluent Heun function**

The Schrödinger equations defining the semiclassical time-dependent two-state problem are written as a system of coupled first-order differential equations for probability amplitudes of the two states $a_{1,2}(t)$ including two arbitrary real functions of time, $U(t)$ (the Rabi frequency, $U>0$) and $\delta(t)$ (detuning modulation, the derivative of which $\delta_t = d\delta/dt$ is the frequency detuning):

$$i\frac{da_1}{dt} = Ue^{-i\delta}a_2, \quad i\frac{da_2}{dt} = Ue^{+i\delta}a_1, \tag{1}$$

which is equivalent to the following linear second-order ordinary differential equation:

$$\frac{d^2 a_2}{dt^2} + \left(-i\delta_t - \frac{U_t}{U}\right)\frac{da_2}{dt} + U^2 a_2 = 0. \tag{2}$$

Here and below the lowercase Latin index denotes differentiation with respect to the corresponding variable.

To find the field configurations for which the solution of Eq. (2) is written in terms of the confluent Heun equation we use the class property of the solvable two-state models [19-21]. According to this property, if the function $a_2^*(z)$ is a solution of this equation rewritten for an auxiliary argument $z$ for some functions $U^*(z)$ and $\delta^*(z)$, then the function $a_2(t) = a_2^*(z(t))$ is the solution of Eq. (2) for the field-configuration defined as

$$U(t) = U^*(z)\frac{dz}{dt}, \quad \delta_t(t) = \delta_z^*(z)\frac{dz}{dt} \tag{3}$$

for an arbitrary complex-valued function $z(t)$. The functions $U^*(z)$ and $\delta_z^*(z)$ are referred to as the amplitude- and detuning-modulation functions, respectively, and the pair $\{U^*, \delta_z^*\}$ is referred to as a basic integrable model. Eqs. (3) together with the transformation of the independent variable $a_2 = \varphi(z)\,u(z)$ reduce Eq. (2) to the following equation for the new dependent variable $u(z)$:

$$u_{zz} + \left(2\frac{\varphi_z}{\varphi} - i\delta_z^* - \frac{U_z^*}{U^*}\right)u_z + \left(\frac{\varphi_{zz}}{\varphi} + \left(-i\delta_z^* - \frac{U_z^*}{U^*}\right)\frac{\varphi_z}{\varphi} + U^{*2}\right)u = 0. \qquad (4)$$

This equation is the confluent Heun equation [37]

$$u_{zz} + \left(\frac{\gamma}{z} + \frac{\delta}{z-1} + \varepsilon\right)u_z + \frac{\alpha z - q}{z(z-1)}u = 0, \qquad (5)$$

when

$$\frac{\gamma}{z} + \frac{\delta}{z-1} + \varepsilon = 2\frac{\varphi_z}{\varphi} - i\delta_z^* - \frac{U_z^*}{U^*} \qquad (6)$$

and

$$\frac{\alpha z - q}{z(z-1)} = \frac{\varphi_{zz}}{\varphi} + \left(-i\delta_z^* - \frac{U_z^*}{U^*}\right)\frac{\varphi_z}{\varphi} + U^{*2}. \qquad (7)$$

Though the general solution of the obtained equations is not known, many particular solutions can be found starting from specific forms of the involved functions. Based on our previous experience, we search for solutions in the following form:

$$\varphi = e^{\alpha_0 z} z^{\alpha_1}(z-1)^{\alpha_2}, \qquad (8)$$

$$U^* = U_0^* z^{k_1}(z-1)^{k_2}, \qquad (9)$$

$$\delta_z^* = \delta_0 + \frac{\delta_1}{z} + \frac{\delta_2}{z-1}. \qquad (10)$$

Then, multiplying Eq. (7) by $z^2(z-1)^2$, we note that for arbitrary $\delta_{0,1,2}$ the product $U_0^{*2} z^{2k_1+2}(z-1)^{2k_2+2}$ is a polynomial in $z$ of maximum fourth degree. Hence, $k_{1,2}$ are integers or half-integers obeying the inequalities $-1 \leq k_{1,2} \cup k_1 + k_2 \leq 0$. This leads to 15 admissible pairs $\{k_1, k_2\}$ defining fifteen classes of models solvable in terms of the confluent Heun function. The corresponding basic models are explicitly presented in Table 1, and according to the class property of integrable models, the actual field configurations $\{U(t), \delta(t)\}$ are given as

$$U(t) = U_0^* z^{k_1}(z-1)^{k_2} \frac{dz}{dt}, \qquad (11)$$

$$\delta_t(t) = \left(\delta_0 + \frac{\delta_1}{z} + \frac{\delta_2}{z-1}\right)\frac{dz}{dt}, \qquad (12)$$

where the involved parameters are generally complex constants which should be chosen so that the functions $U(t)$ and $\delta(t)$ are real for the chosen complex-valued $z(t)$.

Since the parameters $U_0^*$ and $\delta_{0,1,2}$ are arbitrary, all the derived classes are 4-parametric. These classes generalize all the known 3- or 2-parametric subclasses of models solvable in terms of hypergeometric or confluent hypergeometric functions. The classes which extend the 3-parametric hypergeometric classes [20-21] are indicated in Table 1 by "$_2F_1$" and the ones generalizing the 3-parametric confluent hypergeometric classes [19] are marked by "$_1F_1$". Note that the basic models $U^*/U_0^* = 1/z$ and $1/(z-1)$, include both a 3-parametric subclass solvable in terms of $_2F_1$ (the case with $\delta_0 = 0$) and a 3-parametric subclass solvable in terms of $_1F_1$ (the case with $\delta_2 = 0$). 3-parametric subclasses of the classes $U^*/U_0^* = 1/\sqrt{z}$ and $1/\sqrt{z-1}$ specified by the choice $\delta_0 = 0$, $\delta_{1,2} \neq 0$, the solution for which is not written in terms of the hypergeometric functions, was recently presented in [41,42]. The classes $U^*/U_0^* = 1/(\sqrt{z}(z-1))$ and $\sqrt{z}/(z-1)$ extend the 2-parametric subclasses for which the solution is written in terms of the Kummer confluent hypergeometric function [19], and the class $U^*/U_0^* = \sqrt{z/(z-1)}$ includes 2-parametric subclass the solution for which is written in terms of the Gauss hypergeometric function [20].

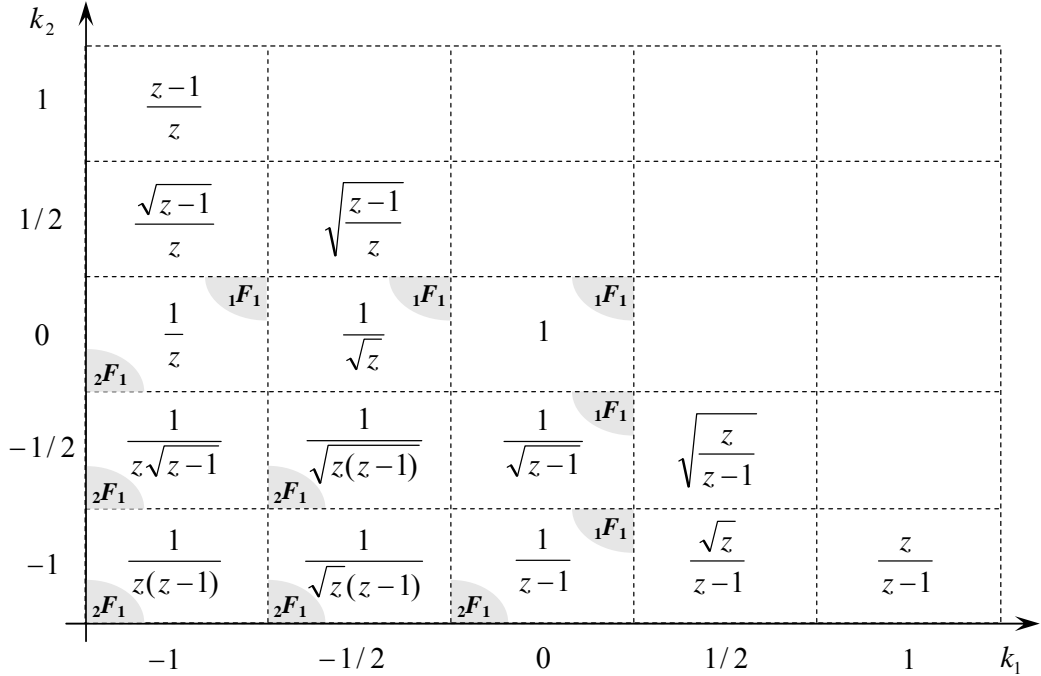

Table 1. Fifteen basic models of amplitude modulation function $U^*$ for which the two-state problem is solved in terms of the confluent Heun functions. The models that include 3-parametric subclasses with $\delta_0 = 0$ solvable in terms of hypergeometric and confluent hypergeometric functions are indicated by "$_2F_1$" and "$_1F_1$", respectively.

## 3. Series expansions of the confluent Heun function

The solution of the initial two-state problem is explicitly written as

$$a_2 = e^{\alpha_0 z} z^{\alpha_1} (z-1)^{\alpha_2} H_C(\gamma, \delta, \varepsilon; \alpha, q; z), \tag{13}$$

where $H_C$ is the confluent Heun function and the parameters $\gamma$, $\delta$, $\varepsilon$, $\alpha$, $q$ are given as

$$\gamma = 2\alpha_1 - i\delta_1 - k_1, \quad \delta = 2\alpha_2 - i\delta_2 - k_2, \quad \varepsilon = 2\alpha_0 - i\delta_0, \tag{14}$$

$$\alpha = -i\delta_0(\alpha_1 + \alpha_2 - \alpha_0) + \alpha_0(\gamma + \delta - \varepsilon) + Q^{(3)}(0)/6, \tag{15}$$

$$q = \alpha_0(\alpha_0 - i\delta_0 - k_1 - i\delta_1) + \alpha_2(1 - \alpha_2 + k_1 + i\delta_1 + k_2 + i\delta_2) +$$
$$\alpha_1(1 - \gamma - \delta + \varepsilon + \alpha_1) - Q''(0)/2 - Q'''(0)/6 \tag{16}$$

with $Q(z) = U_0^{*2} z^{2k_1+2} (z-1)^{2k_2+2}$ and

$$\alpha_0^2 - i\alpha_0\delta_0 = -Q^{(4)}(1)/4!, \quad \alpha_1^2 - \alpha_1(1 + k_1 + i\delta_1) = -Q(0), \quad \alpha_2^2 - \alpha_2(1 + k_2 + i\delta_2) = -Q(1). \tag{17}$$

The confluent Heun function can be constructed via power series expansion:

$$H_C(\gamma, \delta, \varepsilon; \alpha, q; z) = z^\mu \sum_n a_n z^n, \tag{18}$$

where the coefficients obey the following three-term recurrence relation:

$$R_n a_n + Q_{n-1} a_{n-1} + P_{n-2} a_{n-2} = 0 \tag{19}$$

where

$$R_n = (\mu + n)(\mu + n - 1 + \gamma), \tag{20}$$

$$Q_n = q - (\mu + n)(\mu + n - 1 + \gamma + \delta - \varepsilon), \tag{21}$$

$$P_n = -\varepsilon(\mu + n) - \alpha. \tag{22}$$

This series is left-hand side terminated at $n = 0$ if $\mu = 0$ or $\mu = 1 - \gamma$. In general, the convergence radius of the series is equal to unity, however, in some cases the series is right-hand side terminated at some $n = N$ thus producing closed form solutions applicable everywhere. This occurs when $P_N = 0$ and $a_{N+1} = 0$. The latter equation is a polynomial equation of the order $N + 1$ for the accessory parameter $q$ having in general $N + 1$ solutions.

Other expansion functions can be used to construct series solutions, for instance, the hypergeometric and confluent hypergeometric functions [37], Bessel functions [43], Coulomb wave functions [44], incomplete beta functions [45]. For some particular cases, it is possible to use the properties of the derivatives of the solutions of the Heun equation [46], to construct expansions in terms of higher transcendental functions, e.g., the Goursat generalized hypergeometric functions [47]. A particular expansion in terms of the confluent hypergeometric functions, which is convenient for derivation of closed form solutions applicable to the two-state problem discussed here is written as:

$$H_C(\gamma,\delta,\varepsilon;\alpha,q;z) = \sum_{n=0}^{\infty} a_n \cdot {}_1F_1(\alpha_0 + n;\gamma;-\varepsilon z), \qquad (23)$$

where the coefficients of the expansion are given by the recurrence relation

$$R_n a_n + Q_{n-1} a_{n-1} + P_{n-2} a_{n-2} = 0 \qquad (24)$$

with
$$R_n = (n + \alpha_0 - \gamma)(n + \alpha_0 - \alpha/\varepsilon), \qquad (25)$$

$$Q_n = (\gamma - 2(n + \alpha_0))(n + \alpha_0 - \alpha/\varepsilon) + (n + \alpha_0)(\varepsilon - \delta) - q, \qquad (26)$$

$$P_n = (n + \alpha_0)(n + \alpha_0 + \delta - \alpha/\varepsilon), \qquad (27)$$

where $\alpha_0 = \alpha/\varepsilon$ or $\alpha_0 = \gamma$. The series is right-hand side terminated for some non-negative integer $N$ if $P_N = 0$ and $a_{N+1} = 0$. If $\alpha_0 = \alpha/\varepsilon$, the condition $P_N = 0$ is satisfied if

$$\alpha/\varepsilon = -N \quad \text{or} \quad \delta = -N. \qquad (28)$$

If $\alpha_0 = \gamma$, the only choice, since $\gamma$ should not be a negative integer number, is

$$\gamma + \delta - \alpha/\varepsilon = -N. \qquad (29)$$

The termination occurs for $N+1$ values of the accessory parameter $q$ defined from the equation $a_{N+1} = 0$ (or, equivalently, $Q_N a_N + P_{N-1} a_{N-1} = 0$).

## 4. Level-crossing field configurations

Analyzing the derived classes in the general case of variable amplitude- and detuning-modulation functions we note that the most notable feature provided by the extension of the solvable models to the cases covered by the confluent Heun functions is due to the extension of the detuning modulation function to the three-parametric case, which is mathematically manifested by the additional term $\delta_0$ in Eq. (12). Though simple at first glance, this term provides significantly larger range of physically interesting detuning functions as compared with the cases when the problem is solved in terms of the Gauss hypergeometric and the Kummer confluent hypergeometric functions. In the case of constant detuning, $\delta_t(t) = \Delta$, this leads to one or two-peak symmetric or asymmetric pulses with controllable width, among which rectangular box pulses of given width and infinitely narrow pulses are possible as limiting cases. We now will see that by different choices of the independent variable transformation $z = z(t)$ (both real and complex-valued) many other interesting field configurations are modeled by the derived classes when $\delta_t(t) \neq \text{const}$ including symmetric or asymmetric chirped pulses, level-glancing configurations, double and periodically repeated crossings of the resonance.

First of all, a notable point is that due to the $\delta_0$-term the detuning modulation function now allows resonance-crossings at two time-points. This can be easily understood by rewriting $\delta_z^*$ in the following form:

$$\delta_z^* = \frac{\delta_0 z^2 + (-\delta_0 + \delta_1 + \delta_2)z - \delta_1}{z(z-1)}. \tag{30}$$

Now, it is seen that the laser frequency detuning function turns into zero at the points

$$z_{1,2} = \frac{(\delta_0 - \delta_1 - \delta_2) \pm \sqrt{(-\delta_0 + \delta_1 + \delta_2)^2 + 4\delta_0\delta_1}}{2\delta_0} \tag{31}$$

if the discriminant $D = (-\delta_0 + \delta_1 + \delta_2)^2 + 4\delta_0\delta_1 \geq 0$. Hence, if $z_{1,2}$ are inner points of the allowed variation range of $z = z(t)$ and $z'(t) \neq 0$ everywhere within this range, Eq. (12) defines a detuning with *two* crossings of the resonance if $z_1 \neq z_2$ and a configuration *touching* the resonance if $z_1 = z_2$. Otherwise, we have a *non-crossing* model if none of the roots (31) belongs to the variation range of $z(t)$ and a *chirped* pulse if only one root is an inner point of the variation range of $z(t)$. Alternatively, a chirped field configuration is obtained if the numerator in Eq. (30) is canceled to a linear function. This happens if $\delta_0 = 0$ or $\delta_1 = 0$ or $\delta_2 = 0$. Particularly, it is the case for all the three-parametric hypergeometric or confluent hypergeometric models. Since the non-crossing and chirped models have been intensively studied in the past and are well presented in literature we do not discuss those cases here. Instead, we present some examples of field configurations with two resonance-crossings and a specific model describing periodically repeated crossings.

An example of a field configuration with up to two crossings is the one discussed in our recent paper [16]. This is the model, referred to as the generalized Rosen-Zener model since it includes the original non-crossing Rosen-Zener model [48] as a particular constant-detuning case given by the following field configuration:

$$U(t) = U_0 \operatorname{sech}(t), \quad \delta_t(t) = \Delta_0 + \Delta_1 (\operatorname{sech}(t))^2. \tag{32}$$

This model is a member of the class $k_{1,2} = -1/2$ ($U^*/U_0^* = 1/\sqrt{z(z-1)}$) obtained by the real transformation $z = (1 + \tanh(t/\tau))/2$ and the specifications $\delta_0 = 2\Delta_1\tau$, $\delta_1 = -\delta_2 = \Delta_0\tau/2$, $U_0^* = iU_0\tau$, $\tau = 1$. As it is seen, the crossings are due to the parameter $\delta_0$. If $\Delta_1 = -\Delta_0$, the detuning does not actually cross the resonance, but touches it at $t = 0$, thus representing the case of *level-glancing* field configuration [15]. In the general case of two crossings the detuning mimics the parabolic crossing model [17], however, suggesting a finite variation of the detuning instead of the diverging one implied by the exact parabolic model. Note that with the same specifications for $z(t)$ and $\delta_{0,1,2}$, hence, with the same detuning (32), the basic model $k_{1,2} = -1$ produces amplitude modulation $U(t) = U_0 (\text{sech}(t))^2$, and in the case $k_{1,2} = 0$ we have a constant Rabi frequency case: $U(t) = U_0$.

In general, the detuning functions are asymmetric. For instance, this is the case for the transformation $z = (1 + \tanh(t/\tau))/2$ if $\delta_1 \neq -\delta_2$ (for simplicity, we put $\tau = 1$):

$$\delta_t(t) = (\delta_1 - \delta_2) - (\delta_1 + \delta_2)\tanh(t) + \delta_0 (\text{sech}(t))^2 / 2. \tag{33}$$

In this case the parameters $\delta_1$ and $\delta_2$ define the asymptotes of the detuning at $t \to \mp\infty$ that are now not equal (see Fig. 1a).

Moreover, the detuning function may be asymmetric even if $\delta_t(-\infty) = \delta_t(+\infty)$ when an asymmetric transformation $z(t)$ is applied. Such a situation arises, e.g., when the constant amplitude field configuration is considered for the classes for which the function $U^*(z)$ is not symmetric with respect to interchange $k_1 \leftrightarrow k_2$. This is shown in Fig. 1b for the class $k_{1,2} = \{-1, +1\}$. Here, the transformation $z = -W(e^{1-t/\tau})$ (for simplicity, we put $\tau = 1$) with $W$ being the Lambert function [49] produces

$$U(t) = U_0, \quad \delta_t(t) = \delta_0 - \frac{\delta_0 + \delta_1 + \delta_2}{1 + W(e^{1-t})} + \frac{\delta_2}{\left(1 + W(e^{1-t})\right)^2}. \tag{34}$$

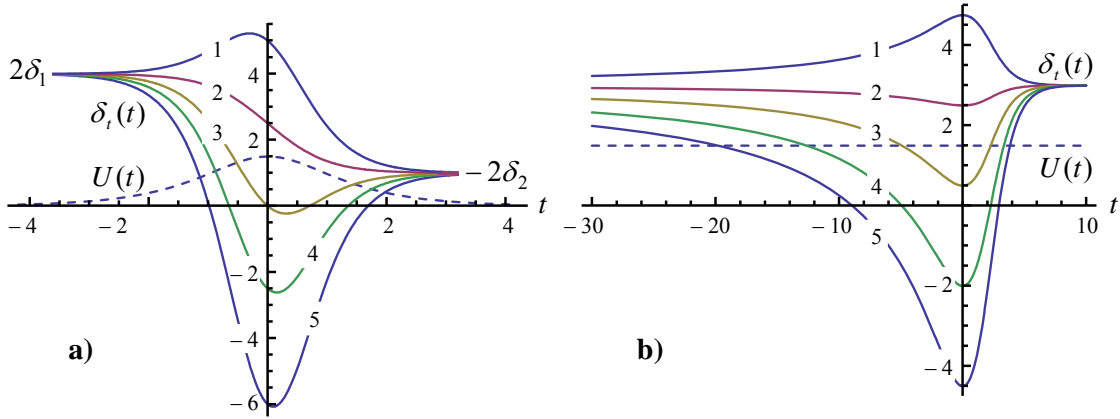

Fig. 1. Asymmetric detuning. All the parameters are assumed dimensionless. Left figure: $k_{1,2} = \{-1/2; -1/2\}$, $U = U_0 \text{sech}(t)$, $U_0 = 1.5$, detuning is given by Eq. (33) with $\delta_1 = 2$, $\delta_2 = -1/2$, $\delta_0 = \{5; 0; -5; -10; -17\}$ (curves 1,2,3,4,5, respectively). Right figure: $k_{1,2} = \{-1, +1\}$, constant Rabi frequency case: $U = U_0 = 1.5$, detuning is given by Eq. (34) with $\delta_0 = 3$, $\delta_1 = -3$, $\delta_2 = \{-7; 2; 10; 20; 30\}$ (curves 1,2,3,4,5, respectively).

Finally, we complete this section by presenting an example of a constant amplitude model with periodically repeated resonance-crossings:

$$U(t) = U_0, \quad \delta_t(t) = A_0 \cos(\Delta t), \quad z(t) = (1 + \sin(\Delta t))/2. \qquad (35)$$

This is a member of the class $k_{1,2} = -1/2$ with the parameters chosen as $U_0^* = iU_0/\Delta$, $\delta_0 = 2A_0/\Delta$, $\delta_{1,2} = 0$. Note that the amplitude of the detuning modulation, that is the maximum deviation $A_0$ of the detuning from the resonance $\delta_t = 0$, is controlled solely by $\delta_0$. This shows once more the usefulness of this parameter.

## 5. Generalized double level-crossing Lorentzian model

Models with two crossings of the resonance are generated also by the complex-valued transformation $z = (1 + i\, y(t))/2$. An example due to the simplest choice $y(t) = t$ is the model referred to as the generalized Lorentzian model which describes the excitation of a two-level atom by a pulse of Lorentzian shape:

$$U(t) = \frac{U_0}{1+t^2}, \quad \delta_t(t) = \Delta_0 + \frac{\Delta_1}{1+t^2}. \qquad (36)$$

This configuration is generated by the basic model $k_{1,2} = -1$ with the specification $\delta_0 = -2i\Delta_0$, $\delta_1 = -\delta_2 = -i\Delta_1/2$ and $U_0^* = iU_0/2$. In this case the crossings are due to the parameters $\delta_{1,2}$. Note that if $k_{1,2} = -1/2$ is considered instead of $k_{1,2} = -1$, we will have $U(t) = U_0/\sqrt{1+t^2}$, and $k_{1,2} = 0$ produces the constant Rabi frequency case $U(t) = U_0$. The time-variation of the detuning (36) for several values of the parameter $\Delta_1$ is shown in Fig. 2.

We have a family of symmetric detuning functions describing both non-crossing and crossing processes with one or two resonance crossing points. If $-\Delta_0/\Delta_1 < 0$ or $-\Delta_0/\Delta_1 > 1$, the detuning does not cross zero. If $\Delta_0 + \Delta_1 = 0$, the detuning touches the origin at $t = 0$ so that in this case we have an model of level-glancing [15]. If $0 < -\Delta_0/\Delta_1 < 1$, there are two crossing points located at $t = \pm\sqrt{-\Delta_1/\Delta_0 - 1}$. In this case the detuning again mimics the parabolic crossing model [17] as the generalized Rosen-Zener configuration (32). However, here also the variation of the detuning is within a finite region, not infinite as it is in the case of the exact parabolic model. Finally, note that $\Delta_0 = 0$ is a specific case when the detuning asymptotically goes to zero as $t \to \pm\infty$.

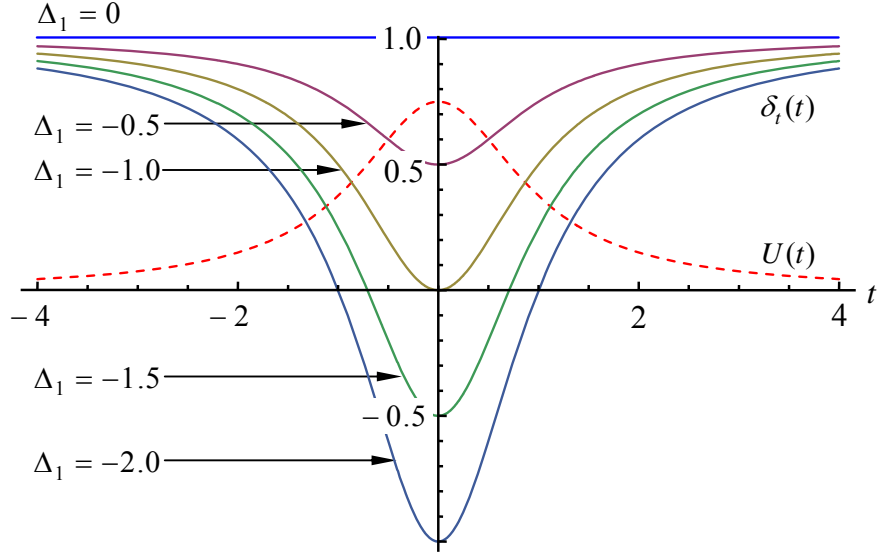

Fig. 2. Generalized double level-crossing Lorentzian model defined by Eqs. (36):
$U_0 = 0.75$, $\Delta_0 = 1$. If $-\Delta_0/\Delta_1 < 0$ or $-\Delta_0/\Delta_1 > 1$, the detuning does not cross zero. If $\Delta_0 + \Delta_1 = 0$, it touches the origin. If $0 < -\Delta_0/\Delta_1 < 1$, there are two crossing points located at $t = \pm\sqrt{-\Delta_1/\Delta_0 - 1}$.

Though the above two symmetric detuning models, generalized level-crossing Rosen-Zener and Lorentzian ones, seem to offer similar double resonance-crossing field configurations, however the close inspection reveals that the very crossing processes have some qualitative differences. These are generated by the path that the variable $z$ draws on the complex plane. In the first case $z$ goes from 0 to 1 along the real axis so that $z$ always belongs to the interval $(0,1)$ connecting two regular singularities of the confluent Heun equation. In contrast to this, in the second case (the Lorentzian model) $z$ changes along the line passing through the midpoint of the interval $z \in (0,1)$ and goes parallel to the imaginary axis starting at $t = -\infty$ from irregular singular point $z = \infty$ of the Heun equation and returns back to the same irregular singularity at $t \to +\infty$. This is a rather complicated case since the behavior of the system is mostly governed by the irregular singularity, though the evolution of the system is strongly influenced by regular singularities $z = 0$ and $z = 1$.

However, fortunately, it turns out that in the case of generalized Lorentzian model again, as it was in the case of constant detuning Lorentzian model, the solution of the problem involves a confluent Heun function with real parameters only. This allows one to get closed form solutions for some sets of the involved parameters using the above series solutions. Indeed, the solution (13) this reads

$$a_2 = z^{\alpha_1}(z-1)^{-\alpha_1} H_C(1+R, 1-R, -2\Delta_0; 0, -(R+\Delta_1/2)\Delta_0; z), \quad z = (1+it)/2, \quad (37)$$

where $R = \sqrt{U_0^2 + \Delta_1^2/4}$ is the effective Rabi frequency and $\alpha_1 = (\Delta_1 + 2R)/4$. Accordingly, the series (23)-(27) may now terminate if the effective Rabi frequency is a natural number. The second termination condition then defines a relation between $\Delta_0$ and $\Delta_1$ for which the termination occurs:

$$\begin{aligned} R=1: \quad & \Delta_0 = 0, \\ R=2: \quad & \Delta_0 = \{0, -4/\Delta_1\}, \\ R=3: \quad & \Delta_0 = \left\{0, -6/(\Delta_1 \pm \sqrt{3+\Delta_1^2/4})\right\}, \ldots \end{aligned} \quad (38)$$

For the lowest order termination $\{R, \Delta_0\} = \{1, 0\}$ the solution of the two-state problem (1) is explicitly written as

$$a_2 = e^{i(1+\Delta_1/2)\arctan(t)} \left( C_1 + \frac{C_2}{1+it} \right), \quad (39)$$

and for the first non-trivial case $\{R, \Delta_0\} = \{2, -4/\Delta_1\}$ we have

$$a_2 = \left( \frac{1-it}{1+it} \right)^{-\Delta_1/4} \left( C_1 \frac{8 + 4i\Delta_1 t + 8t^2 + \Delta_1^2}{1+t^2} + C_2 \frac{e^{-4it/\Delta_1}}{1+t^2} \right). \quad (40)$$

If the system starts from the first state, i.e., the initial condition $a_2(-\infty) = 0$ is considered, $C_1$ becomes zero and only the second term remains. It is checked that then $a_2$ is zero also at $t = +\infty$, so that in this case the system returns to its initial state at the end of the interaction.

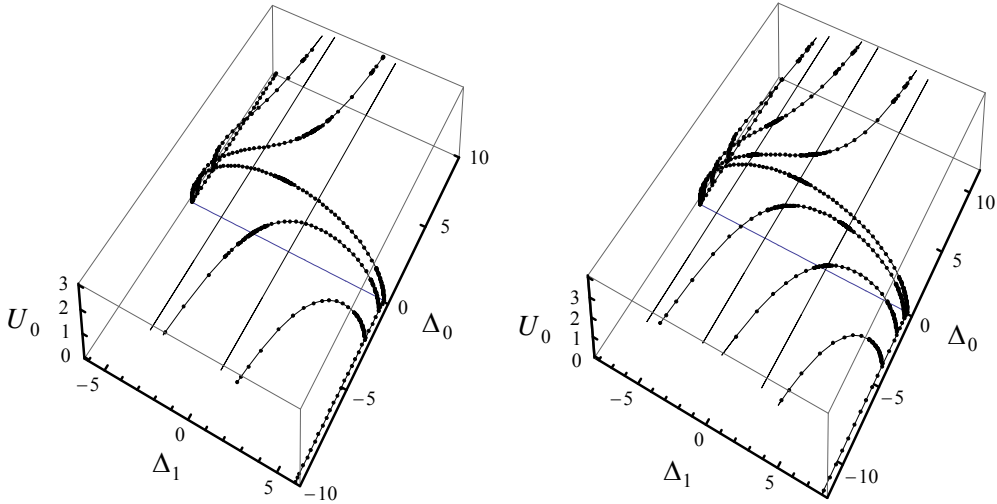

Fig. 3. Curves defined by Eqs. (38) for $R = 3$ and $R = 4$.

The same feature, i.e. the return of the system to its initial state at the end of the interaction if the system starts from the first level, is observed for all subsequent sets $\{R,\Delta_0\}$ given by Eqs. (38). Thus, the equations (38) define curves in the 3D space of the parameters $\{U_0,\Delta_0,\Delta_1\}$ belonging to the complete return spectrum of the system. The curves for $R=2$ and $R=3$ are shown in Fig. 3.

## 6. Summary

Thus, we have discussed 15 *four*-parametric classes of time dependent two-state models allowing solution of the problem in terms of the confluent Heun functions. These classes extend all the known three- and two-parametric classes of two-state models solvable in terms of the Gauss hypergeometric functions (there are six such classes) and in terms of the Kummer confluent hypergeometric functions (five classes) to a more general type of the detuning modulation function including an additional parameter. It should be noted that there are very few papers discussing the solutions of the two-state problem in terms of the Heun functions. The biconfluent Heun equation was applied in [50] to generalize the models solvable in terms of the confluent hypergeometric functions and the general Heun equation was used in [51] to study the two-state problem for an atom interacting with the bichromatic field of two lasers. Besides, three 2-parametric families of pulses for which, however, the involved confluent Heun functions are degenerated to the hypergeometric or confluent hypergeometric functions are presented in [19,20]. Finally, two 3-parametric families belonging to the classes $k_{1,2}=\{-1/2,0\}$ and $\{0,-1/2\}$ are discussed in [41,42]. In the latter cases, however, only the detuning modulation functions with $\delta_0=0$ are discussed. This is a rather restrictive condition because it is this parameter that allows generation of a considerably wider variety of field configurations as compared with the known hypergeometric classes. For instance, this parameter controls the pulse width in the constant detuning case, and it is this parameter that leads to double and periodically repeated crossings of the resonance in the variable detuning case.

Analyzing the derived classes in the general case of variable Rabi frequency and detuning functions we note that the most notable feature provided by the extension of the solvable models to the cases covered by the confluent Heun functions is due to an extra constant term in the detuning modulation function. Though simple at first glance, this term provides significantly larger range of physically interesting detuning functions as compared with the cases when the problem is solved in terms of hypergeometric functions. Due to this term the derived classes provide numerous models with two resonance-crossing time points. Such models are generated by both real and complex transformations of the independent variable. In general the resulting detuning functions are asymmetric, the asymmetry being controlled by the parameters of the detuning modulation function. For some classes, however, the asymmetry may be additionally caused by the amplitude modulation function. We have presented an example of the latter possibility. Furthermore, we have mentioned a constant amplitude model with periodically repeated resonance-crossings which is again due to the extra constant term in the detuning modulation function.

Finally, we have discussed an interesting field configuration, a member of the class $k_{1,2} = -1$ generated by the complex-valued transformation $z = (1 + it)/2$, which describes the excitation of a two-level atom by a pulse of Lorentzian shape. This model referred to as the generalized Lorentzian model suggests a family of symmetric or asymmetric detuning functions describing both non-crossing and crossing processes with one or two crossings of the resonance. In the case of one crossing point this is an example of level-glancing. We used a particular series expansion of the solution of the confluent Heun equation in terms of the Kummer hypergeometric functions to derive particular closed form solutions of the two-state problem for this field configuration both for the constant and variable detuning cases. The particular sets of the involved parameters for which these closed form solutions are obtained turned to define curves in the 3D space of the involved parameters belonging to the complete return spectrum of the considered two-state quantum system.

**Acknowledgments**

This research has been conducted within the scope of the International Associated Laboratory (CNRS-France & SCS-Armenia) IRMAS. The research has received funding from the European Union Seventh Framework Programme (FP7/2007-2013) under grant agreement No. 205025 – IPERA.